\documentclass[10pt,journal,twoside,web]{ieeecolor}
\usepackage{generic}

\usepackage[colorlinks,linkcolor=blue,anchorcolor=blue,citecolor=blue]{hyperref}

\usepackage{amsmath}
\usepackage{multicol}
\usepackage{autobreak}
\usepackage{mathrsfs} 
\usepackage{amsfonts,amssymb} 
\usepackage{microtype} 

\usepackage{amssymb} 
\usepackage{easyReview} 
\usepackage{cuted} 
\usepackage{bm} 
\usepackage{tabu}

\usepackage{subfigure}
\usepackage{makecell}
\usepackage{threeparttable} 
\usepackage{booktabs}
\usepackage{arydshln} 
\usepackage{enumerate} 
\usepackage{multirow} 

\usepackage{soul} 
\setuldepth{a}
\usepackage{bbding}

\def\BibTeX{{\rm B\kern-.05em{\sc i\kern-.025em b}\kern-.08em
    T\kern-.1667em\lower.7ex\hbox{E}\kern-.125emX}}
\begin{document}
\title{Multiscale Fusion for Abnormality Detection and Localization of Distributed Parameter Systems}

\author{Peng~Wei, and Han-Xiong~Li,~\IEEEmembership{Fellow,~IEEE} 
	\thanks{This work was supported in part by a GRF project from RGC of Hong Kong under Grant CityU: 11206623 and in part by a project from City University of Hong Kong under Grant 7005680. \textit{(Corresponding author: Han-Xiong~Li.)}}
	\thanks{Peng Wei is with the Department of Systems Engineering, City University of Hong Kong, Hong Kong, China, and also with the School of Automation, Wuhan University of Technology, Wuhan 430070, China (e-mail: pengwei7-c@my.cityu.edu.hk).}
	\thanks{Han-Xiong Li is with the Department of Systems Engineering, City University of Hong Kong, Hong Kong, China (e-mail: mehxli@cityu.edu.hk).}
}

\maketitle

\begin{abstract}
Numerous industrial thermal processes and fluid processes can be described by distributed parameter systems (DPSs), wherein many process parameters and variables vary in space and time. Early internal abnormalities in the DPS may develop into uncontrollable thermal failures, causing serious safety incidents. In this study, the multiscale information fusion is proposed for internal abnormality detection and localization of DPSs under different scenarios. We introduce the dissimilarity statistic as a means to identify anomalies for lumped variables, whereas spatial and temporal statistic measures are presented for the anomaly detection for distributed variables. Through appropriate parameter optimization, these statistic functions are integrated into the comprehensive multiscale detection index, which outperforms traditional single-scale detection methods. The proposed multiscale statistic has good physical interpretability from the system disorder degree. Experiments on the internal short circuit (ISC) of a battery system have demonstrated that our proposed method can swiftly identify ISC abnormalities and accurately pinpoint problematic battery cells under various working conditions.

\end{abstract}


\section{Introduction}
Numerous industrial processes, such as chemical reactions, heat exchangers, and fluid dynamics systems, can be described by distributed parameter systems (DPSs) \cite{christofides2002nonlinear}, wherein the input, output, and process variables vary spatially and temporally \cite{chen2021modified}. Early internal abnormalities in the DPS, if not promptly identified and addressed, may evolve into uncontrollable failures \cite{feng2016online}. Such failures can escalate, leading to serious safety incidents that not only disrupt the process but also pose significant risks to personnel and equipment. Therefore, understanding and monitoring these early signs are critical in preventing catastrophic outcomes in these complex industrial systems.


Traditional methods for detecting abnormalities in DPSs can be categorized into first-principle methods \cite{feng2018detection,wang2019model,xiong2019online} and data-based approaches \cite{feng2022online,jiang2021data}. For the first-principle methods, the mathematical models of process variables such as temperature, flow, and air pressure should be derived based on governing equations. Then, the measured values of the process variables will be compared with their model-predicted ones. If the residual exceeds the preset threshold, an abnormality is considered to have occurred. For example, based on the dual-Kalman filter, an open-circuit voltage-based diagnostic model was proposed in \cite{ma2021fault} for the external soft-short circuit for series-connected battery packs. However, industrial processes are usually governed by many complex partial differential equations (PDEs) \cite{deng2021reduced,xia2017computationally}, which are difficult to derive accurately in practice.   

Conventional data-based methods utilize sensor measurements to detect abnormalities in the DPS without relying on governing equations \cite{wei2021spatial,ojo2020neural}. For instance, a method for ISC detection \cite{schmid2022early} was introduced for series-connected battery packs. This method relies on nonlinear process monitoring of the voltage differences among the cells within the battery pack. In the context of state representation methods, Jiang et al.\cite{jiang2021data} utilized normalized cell voltage for early fault diagnosis of battery packs. These methods do not consider the effect of temperature variation since it changes very slowly in the early stage of failure. However, the temperature distribution is sensitive to thermal abnormalities and may be helpful in fault diagnosis of DPSs.

After abnormalities are detected, it is necessary to locate the abnormal units in the DPS so that they can be replaced in time. Taking the battery system as an example, abnormal battery cells need to be replaced in time to reduce the risk of failure of the entire battery system. For the battery system, the abnormality localization algorithms mainly rely on the voltages of battery cells \cite{xu2023online,schmid2020data,chen2022joint}. For example, Schmid et al. proposed a data-driven fault diagnosis method \cite{schmid2020data} for battery systems using cross-cell voltages. These methods have good performance in series circuits. However, they are not effective for abnormality localization in parallel circuits since the voltage signal of each cell is the same in this situation.

Process variables of a DPS can be categorized into lumped and distributed parameters. Distributed parameters, exemplified by temperature, are spatiotemporal variables that exhibit variations both spatially and temporally, reflecting the complex dynamics of the system. Conversely, lumped parameters, such as voltage and current, are one-dimensional variables, characterized by their dependence solely on either space or time. Current fault diagnosis methods mainly focus on distributed parameters or centralized parameters. However, for the entire DPS, faults may cause abnormalities in distributed parameters or lumped parameters.
Currently, there is no method that considers both distributed and lumped parameters to design comprehensive statistics for fault detection and localization. Multiscale fault diagnosis methods can monitor the system from different angles, thereby improving algorithm robustness and system stability.


Based on the above considerations, the multiscale information fusion-based abnormality detection and localization framework is proposed for DPSs. First, the dissimilarity statistic is proposed for disorder degree evaluation of lumped parameters, and the spatial statistic and temporal statistic are used for that of distributed parameters. On this basis, the design of multiscale statistics is transformed into an optimization problem, which can be addressed by traditional optimization algorithms. The proposed multiscale data-driven fault diagnosis method can detect and locate the internal abnormalities of DPSs under various conditions. 

The primary contributions of this research can be outlined as follows:
\begin{itemize}
	\item [1)] The multiscale information fusion (MIF) method is proposed to design comprehensive statistics for reliable detection and localization of internal abnormalities in DPSs. 
	\item [2)] The dissimilarity statistic, spatial statistic, and temporal statistic are devised across different scales and subsequently integrated into a comprehensive multiscale statistic.
	\item [3)] Experiments involving internal short circuit (ISC) tests were carried out on a battery system, revealing that the proposed method is capable of accurately detecting abnormal battery cells with a small detection delay.
\end{itemize}

\section{Problem description}
\begin{figure}[htbp] 
	\centering
	\includegraphics[width=0.45\textwidth]{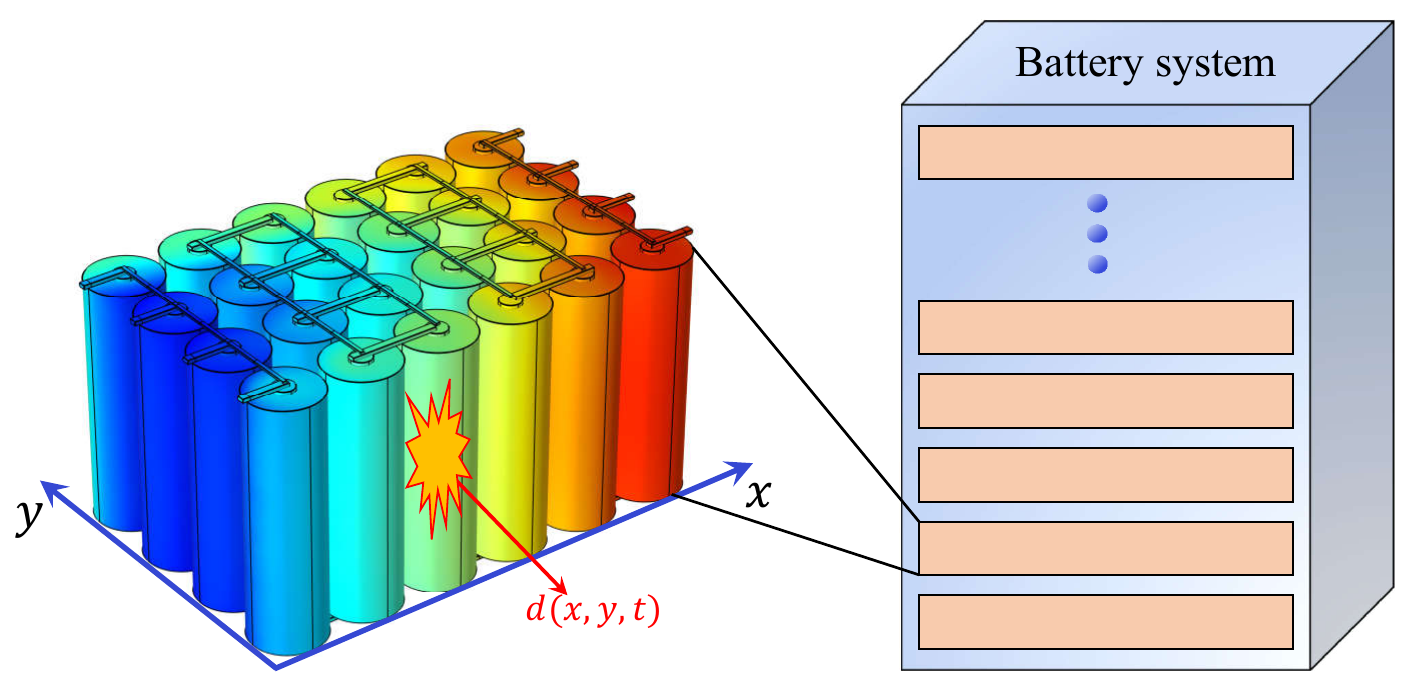}
	\caption{Schematic diagram of a battery pack in the battery system.}
	\label{fig:battery_system}
\end{figure}

As shown in Fig. \ref{fig:battery_system}, the battery system is composed of many battery packs. The thermal behavior of a battery pack can be simplified as a two-dimensional DPS and characterized by the following partial differential equation (PDE):
\begin{equation}\label{equ:PDE}
	\frac{\partial T(x,y,t)}{\partial t} = k_x \frac{\partial^2 T}{\partial^2 x} + k_y \frac{\partial^2 T}{\partial^2 y} + d(x,y,t)+ b^T(x,y) u(t) 
\end{equation} 
subject to the following boundary conditions:
\begin{equation*}\label{equ:boundary_condition}
	\left\{\begin{split}
		\left.q_x\left(T, \frac{\partial T}{\partial x}\right)\right|_{x=0 \text { or } x=x_{b}}=0 \\
		\left.q_y\left(T, \frac{\partial T}{\partial y}\right)\right|_{y=0 \text { or } y=y_{b}}=0
	\end{split}\right.
\end{equation*}
and the following initial condition:
\begin{equation*}
	T(x, y, 0)=T_{0}(x,y)
\end{equation*}
\noindent incorporating the thermal generation term $u(t)$ and its corresponding spatial distribution function $b(x, y)$ defined as follows:
\begin{equation}\label{}
	\left\{\begin{array}{l}
		b(x,y) = \left[ b_1(x,y),b_2(x,y),\cdots,b_N(x,y)\right]^T \\
		u(t) = [ u_1(t),u_2(t),\cdots,u_N(t)]^T \\
		u_i(t) = f(V_i(t),I_i(t),\bar{T}_i(t),SOC_i(t))
	\end{array}\right.
\end{equation}
\noindent in which $T(x, y, t)$ represents the temperature variable; $x$ ranges from 0 to $x_b$, $y$ ranges from 0 to $y_b$ for spatial coordinates; and $t$ ranges from 0 to infinity for the temporal variable; $ k_x $ and $ k_y $ are unknown functions along $ x $ and $ y $ directions; $ d(x,y,t) $ represents the unknown abnormalities; $N$ represents the quantity of cells within the battery pack; $q_x(\cdot)$ and $q_y(\cdot)$ denote nonlinear functions corresponding to the unknown mixing boundaries; $T_0(x, y)$ represents the initial output; $u_i(t)$ corresponds to the thermal generation term for the $i$-th battery cell, and it is linked to the spatial distribution function $b_i(x, y)$, where $ i=1,2,\cdots,N $; $ \bar{T}_i(t) $ represents the average temperature of the $ i $-th cell; $ V_i(t) $, $ I_i(t) $, and $ SOC_i(t) $ represent the terminal voltage, load current, and state of charge (SOC) of the $ i $-th cell, respectively.

The abnormality of a single unit can potentially trigger uncontrollable failures in the DPS. For example, the thermal abnormality in a battery cell may lead to thermal runaway throughout the entire battery system. Therefore, it is necessary to promptly detect abnormalities and locate abnormal units within the DPS. However, it is challenging to achieve fast detection and precise localization of the spatiotemporal abnormality source $ d(x,y,t) $ due to the following reasons:
\begin{itemize}
	\item [1)] Traditional first-principle methods depend on the governing PDE (\ref{equ:PDE}) and corresponding boundary conditions, which are challenging to obtain in practical industrial processes.
	\item [2)] Existing data-based methods mainly target a single type of variable (lumped variable or distributed variable), which limits the performance of these methods under various working conditions.
\end{itemize}

\section{Multiscale Information Fusion}
\begin{figure}[htbp] 
	\centering
	\includegraphics[width=0.48\textwidth]{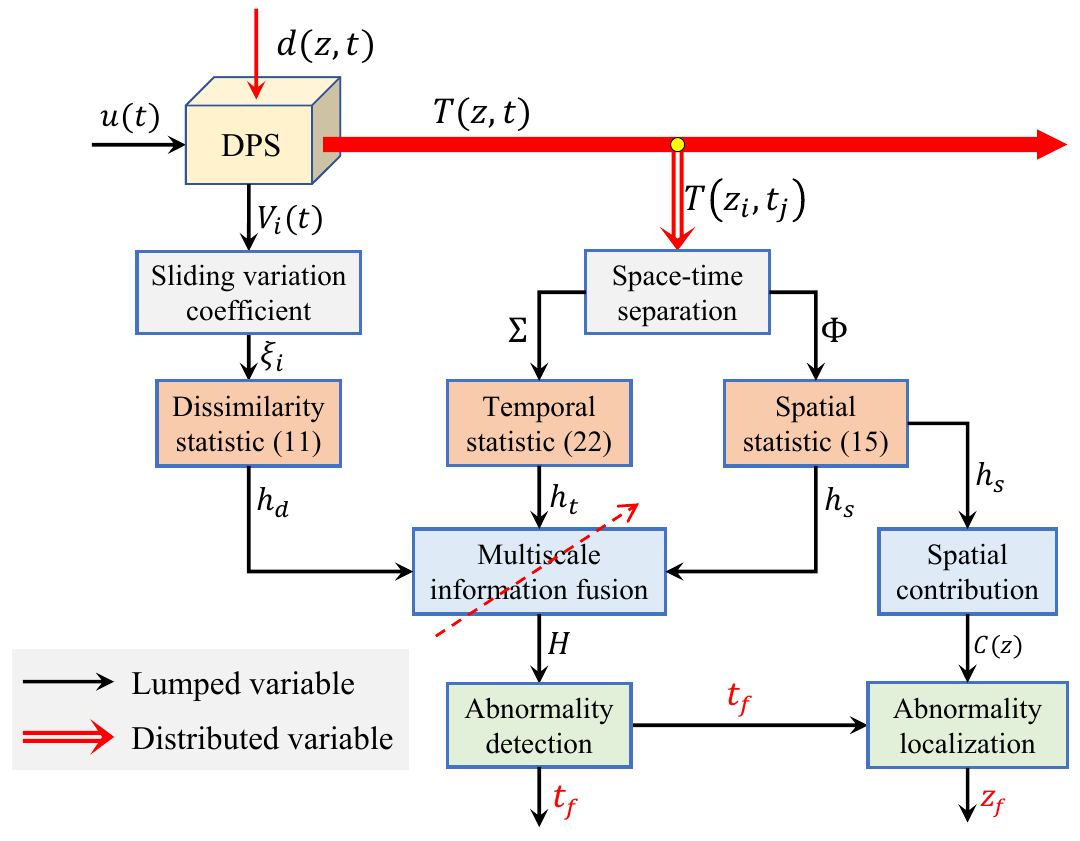}
	\caption{Framework of the proposed multiscale information fusion (MIF).}
	\label{fig:framework}	
\end{figure}

The outputs measured from a DPS can be divided into lumped and distributed variables. Let $ z= [x,y] $. The distributed variable, e.g., the temperature $ T(z,t) $, changes with both space and time, whereas the lumped one, e.g., the cell voltage $ V_i(t) $, vary only with time. There is no spatial coupling between the lumped variables, or the spatial coupling is small enough to be ignored in the same system.  As shown in Fig. \ref{fig:framework}, the multiscale information fusion (MIF) is introduced to detect and identify abnormalities in DPSs by leveraging both lumped and distributed variables.

\subsection{Dissimilarity Statistic for Lumped Variables}
\begin{figure}[htbp] 
	\centering
	\includegraphics[width=0.3\textwidth]{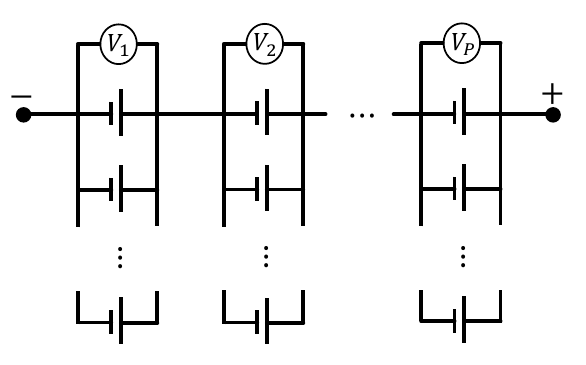}
	\caption{Circuit diagram of a battery pack.}
	\label{fig:circuit_diagram}
\end{figure}

Typical lumped variables for battery systems include voltage and current measured from different battery cells. The dissimilarity statistic is constructed to assess the disorder degree of these lumped variables as follows. Assume that there are $ P $ number of parallel circuits in the battery pack, and each parallel circuit is mounted with a voltmeter, as shown in Fig. \ref{fig:circuit_diagram}. The voltage data $ \{V_i(t)\}_{i=1,t=1}^{P,L} $ is measured from battery cells and used for dissimilarity statistic calculation; $ V_i(t) $ denotes the voltage of the $ i $-th parallel circuit at time $ t $; $ L $ represents the total sampling time length.

\subsubsection{Sliding Variation Coefficient} The variation coefficient is widely used in many fields, including biology, economics, psychology, engineering, and reliability theory, to measure the dispersion of a probability distribution \cite{jalilibal2021monitoring,groeneveld2011influence}. Here, the sliding variation coefficient is proposed to design the dissimilarity statistic for real-time fault detection as follows:
\begin{equation}\label{equ:sliding_variation coefficient}
	\xi_i(k)=\frac{\sigma_i(k)}{\mu_i(k)} 
\end{equation}
where $ \xi_i(k) $ denotes the sliding variation coefficient of the $ i $-th cell voltage at time $ k $; $ \mu_i(k) $ and $ \sigma_i(k) $ are the mean and the standard deviation of the $ i $-th cell voltage at time $ k $, respectively, and calculated as follows
\begin{equation*}
	\mu_i(k) = \frac{\int_{k-W+1}^{k} V_i(t) dt}{W}
\end{equation*}
\begin{equation*}
	\sigma_i(k) = \sqrt{\frac{\int_{k-W+1}^{k} (V_i(t) - \mu_i(k))^2 dt}{W}}
\end{equation*}
with $ W $ denoting the sliding window size.

\subsubsection{Normalization} The Z-score, also called the standard score \cite{chadha2016exploring}, is utilized to normalize the sliding variation coefficient as follows:
\begin{equation}\label{equ:Z_score}
	Z_i(k) = \frac{|\xi_i(k) - \mu_\xi(k)|}{\sigma_\xi(k)}
\end{equation}
where $ Z_i(k) $ denotes the Z-score of the voltage of the $ i $-th parallel circuit at time $ k $; $ \mu_\xi(k) $ and $ \sigma_\xi(k) $ are the mean and the standard deviation of the sliding variation coefficient of all the voltages at time $ k $, i.e., $ \{\xi_i(k)\}_{i=1}^P $.

\subsubsection{Dissimilarity Statistic} 
The skewness, a measure of the asymmetry of the probability distribution of a random variable \cite{doane2011measuring} in statistics, is modified to construct the dissimilarity statistic as follows:
\begin{equation}\label{equ:dissimilarity_statistic}
	h_d(k) = \frac{\frac{1}{P}\sum_{i=1}^{P}|Z_i(k) - \mu_Z(k)|^3}{\left(\frac{1}{P}\sum_{i=1}^{P}(Z_i(k) - \mu_Z(k))^2\right)^{3/2}}
\end{equation}
where $ h_d(k) $ is the dissimilarity statistic of $ \{Z_i(k)\}_{i=1}^P $, which evaluates the disorder degree of the $ P $ numbers of voltage signals from the battery system. $ \mu_Z(k) $ is the mean of $ \{Z_i(k)\}_{i=1}^P $. Under normal conditions, the sliding variation coefficients $ \xi_i(k)_{i=1}^P $ of voltages will be symmetrically distributed on both sides of their mean, resulting in a small dissimilarity statistic. On the contrary, a relatively higher dissimilarity statistic will be obtained under abnormal conditions.

\subsection{Spatial and Temporal Statistic for Distributed Variables}
The temperature output is a typical distributed variable since the temperature $ T(x,y,t) $ is governed by the PDE in (\ref{equ:PDE}). In addition, temperatures measured from different cells would affect each other, i.e., there is a spatial coupling between them. Due to the spatiotemporal coupling characteristics of distributed variables, the spatial statistic and temporal statistic should be derived appropriately to describe the system dynamics along spatial and temporal dimensions. Assume each battery cell is mounted with a temperature sensor. The temperature data $ \{T(z_i,t_j)\}_{i=1,j=1}^{N,L} $ measured from battery cells is used for spatiotemporal statistic calculation; $ z_i = [x_i,y_i]^T$ denotes the spatial coordinate of the $ i $-th battery cell.

Due to the complex couplings of distributed temperature, the space-time separation technique \cite{li2010modeling} is used to extract the spatial and temporal dynamics as follows:
\begin{equation}\label{equ:space_time-separation}
	Y^k = \Phi^k \Lambda^k A^k
\end{equation}
in which $ Y^k \in \mathbb{R}^{N \times W} $ denotes the temperature data matrix composed of $ \{T(z_i,t_j)\}_{i=1,j=k-W+1}^{N,k} $ with $ W $ representing the window size $ (W \leq k) $; $ Y^k(i,j) \triangleq T(z_i,t_j) $; $ \Phi^k \in \mathbb{R}^{N\times n} $ is the spatial basis function (SBF) matrix; $ \Lambda^k \in \mathbb{R}^{n\times n} $ denote the singular value matrix; $ A^k \in \mathbb{R}^{n\times W} $ represent the temporal coefficient matrix; $ n\leq N $ denotes the model order. The detailed derivation of (\ref{equ:space_time-separation}) can be referred to \cite{li2010modeling}.

\subsubsection{Spatial Statistic} The SBFs capture the spatial dynamics of distributed variables. Spatial statistic will be constructed based on the SBF matrix $ \Phi^k $ as follows:
\begin{equation}\label{key}
	\Delta \Phi^k = \sum_{i=1}^n |\phi_i^k - \phi_i^0|
\end{equation}
in which $ \phi_i^k $ is the $ i $-th column of the matrix $ \Phi^k $, that is, the $ i $-th SBF at time point $ k $; The $ i $-th initial SBF, denoted as $ \phi_i^0 $, is obtained from the initial matrix $ Y^0 \triangleq Y^W $ with $Y^W$ consisting of $ \{T(z_i,t_j)\}_{i=1,j=1}^{N,W} $; $ n $ denotes the model order of the system.

The probability density function (PDF) of SBF variation at time point $ k $, denoted as $ p^k(z) $, will be constructed as:
\begin{equation}\label{key}
	p^k(z) \triangleq \Delta \Phi^k (z)/{G}
\end{equation}
where $ G $ is defined as follows:
\begin{equation}\label{key}
	G = \int_{\Gamma} \Delta \Phi^k(z) dz
\end{equation}
in which $ \Gamma $ signifies the length of 1-D space. When dealing with high-dimensional space, it becomes essential to break it down into a combination of several one-dimensional spaces. As for spatially discrete form, $ G $ is defined as $ G \triangleq \sum_{i=1}^{N} \Delta \Phi^k(i) $. 

According to the PDF of SBF variations, we can construct the spatial statistic as follows:
\begin{equation}\label{equ:spatial_statistic}
	h_s(k) = 1+\sum_{i=1}^2 P_i^k \log_2(P_i^k)
\end{equation}
in which $ P_1^k $ and $ P_2^k $ are defined as $ P_1^k=\int_{0}^{\Gamma/2} p^k(z) dz $ and $ P_2^k=\int_{\Gamma/2}^{\Gamma} p^k(z) dz=1-P_1^k $, respectively. 
\subsubsection{Temporal Statistic} The approximate entropy, sample entropy, and fuzzy entropy are three classic measures of complexity of time series \cite{chen2009measuring}. The temporal statistic is constructed by the fuzzy entropy since it is more effective and less sensitive to parameter selection than the other two methods \cite{chen2009measuring}. The temporal coefficients reflect the temporal dynamics of the distributed variables.

Based on the $ i $-th temporal coefficient $ a_i^k $, that is, the $ i $-th row of $ A^k $, a set of vectors can be constructed as:
\begin{equation}\label{key}
	\mathbf{X}_j^{k} = [a_i^k(j),a_i^k(j+1),\cdots,a_i^k(j+m-1)] - u_i^k(j)
\end{equation}
where $ \mathbf{X}_j^{k} \in \mathbb{R}^m $ $ (m \leq N-2) $; $ j=1,2,\cdots,W-m+1 $; $ m $ is a constant and can be selected through cross-validation; $ m=2 $ in this research. $ u_i^k(j) $ is the baseline corresponding to $ i $-th temporal coefficient and defined as:
\begin{equation}\label{key}
	u_i^k(j) = \dfrac{1}{m} \sum_{l=0}^{m-1} a_i^k(j+l)
\end{equation}

Based on the fuzzy membership function $ f(x)=\exp (-\ln(2) (x/r)^2) $, the similarity degree $ D_{jq}^{k} $ between the vector $ \mathbf{X}_j^{k} $ and its neighboring vector $ \mathbf{X}_q^k $ can be defined as:
\begin{equation}\label{key}
	\begin{split}
		D_{jq}^{k} =& f(d_{jq}^{k}) \\ 
		=& \exp \left(-\ln(2) \left(\dfrac{d_{jq}^{k}}{r}\right)^2\right)
	\end{split}
\end{equation}
where $ d_{jq}^k $ is the maximum distance between vectors $ \mathbf{X}_j^{k} $ and $ \mathbf{X}_q^k $, and defined as:
\begin{equation}\label{key}
	\begin{split}
		d_{jq}^{k} = \max_{l=1,2,\cdots,m} &\left(|a_i^k(j+l-1)-u_i^k(j)|\right. \\
		&\left. - |a_i^k(q+l-1)-u_i^k(q)|\right)
	\end{split}	
\end{equation}
The average similarity degree between vectors $ \mathbf{X}_j^{k} \ (j=1,2,\cdots,W-m) $ can be calculated as:
\begin{equation}\label{equ:average_similarity_degree}
	S_i^m(k) = \frac{1}{(W-m)(W-m-1)} \sum_{j =1}^{N-m}\sum_{q=1,q \neq j}^{N-m} D_{jq}^{k}
\end{equation}
Based on (\ref{equ:average_similarity_degree}), the fuzzy statistic of the $ i $-th temporal coefficient can be calculated as:
\begin{equation}\label{key}
	h_f(a_i^k) = \ln S_i^m(k) - \ln S_i^{m+1}(k) 
\end{equation}

Based on the fuzzy statistic of all temporal coefficients, we can construct the temporal statistic as follows:
\begin{equation}\label{equ:temporal_statistic}
	\begin{split}
		h_t(k) =& \sum_{i=1}^{n} \lambda_i^k h_f(a_i^k) \\
		=& \sum_{i=1}^{n} \lambda_i^k (\ln S_i^m(k) - \ln S_i^{m+1}(k))
	\end{split}
\end{equation}
in which $ \lambda_i^k $ is the $ i $-th singular value in (\ref{equ:space_time-separation}), i.e., the $ i $-th diagonal element of the matrix $ \Lambda^k $.

\subsection{Design of Multiscale Statistic} 
For reliable abnormality detection, a general statistic should be formulated to comprehensively assess the level of disorder within a DPS across various scales.
\subsubsection{Multiscale Information Fusion} Assume $h_d(k)$ denote the statistic of lumped variables at time $k$. A multiscale statistic $ H(k) $ can be designed to evaluate the disorder degree of the DPS comprehensively:
\begin{equation}\label{equ:mutiscale_statistic}
	H(k) = \alpha_1 [h_d(k)] + \alpha_2 [h_s(k)] + \alpha_3 [h_t(k)]
\end{equation}
with $ \alpha_1 + \alpha_2 +\alpha_3 = 1 $; $ \alpha_i \in [0,1]$ is a weighting parameter; $ [\cdot] $ is the normalization operator and defined as $ [f(t)] = f(t)/\max(f(t)) $. The maximum values of $ h_d(k) $, $ h_s(k) $, and $ h_t(k) $ can be calculated from the training data $ \{T(z_i,t_j)\}_{i=1,j=1}^{N,L_1} \ (L_1 \leq L)$, in which $ L_1 $ denotes the time length of training data.

\subsubsection{Parameter Optimization} For achieving higher abnormality detection rates, lower false alarming rates, and smaller abnormality detection delay, the window size $ W $, and the weighting parameters $ \alpha_1,\ \alpha_2,\ \alpha_3 $ should be optimized. The optimization objective can be expressed as follows:
\begin{equation}\label{equ:optimization_objective}
	\min_{W,\alpha_1,\alpha_2,\alpha_3} \dfrac{1}{\eta_1} + \eta_2 + \eta_3			
\end{equation}
subject to
\begin{equation}\label{key}
	\left\{\begin{array}{l}
		\alpha_1 + \alpha_2 + \alpha_3 = 1 \\
		\alpha_i \in [0,1] \\
		W \in Z^+,\ W \geq 1
	\end{array}\right.
\end{equation}
where $ \eta_1 $, $ \eta_2 $, and $ \eta_3 $ represent the abnormality detection rate (ADR), false alarming rate (FAR), and relative abnormality detection delay (ADD), respectively, and are defined as follows:
\begin{equation}\label{key}
	\eta_1 = \dfrac{N_{da}}{N_{ta}} \times 100\%
\end{equation}
\begin{equation}\label{key}
	\eta_2 = \frac{N_f}{N_{tn}} \times 100\%
\end{equation}
\begin{equation}\label{key}
	\eta_3 = \frac{t_d - t_a}{t_r}
\end{equation}
where $ N_{da} $ and $ N_f $ represent the count of correctly detected abnormal samples and the count of falsely detected abnormal samples, respectively; $ N_{ta} $ and $ N_{tn} $ represent the overall count of  abnormal samples under abnormal conditions and the overall count of normal samples under normal conditions, respectively; $t_d$ represents the moment when the abnormality is initially detected, while $t_a$ denotes the moment when the abnormality actually occurs; $ t_r $ is a reference time used to balance the weights of the optimization objective (\ref{equ:optimization_objective}), i.e., prevent abnormality detection delay from dominating the entire optimization problem. 

Problem (\ref{equ:optimization_objective}) is essentially a multi-parameter optimization problem that can be effectively addressed using several algorithms, such as the genetic algorithm, the simulated annealing algorithm, and the immune algorithm \cite{amaran2016simulation}. Here, the modified genetic algorithm (MGA) \cite{song2019improved} is employed for parameter optimization in (\ref{equ:optimization_objective}). Different from traditional genetic algorithms, the MGA has better global search ability and faster convergence speed. The fitness function of genetic algorithm is the same as the optimization objective (\ref{equ:optimization_objective}). The training data under normal and abnormal conditions are used to calculate $\eta_1$, $\eta_2$, and $\eta_3$ of the fitness function.

\subsubsection{Physical Interpretability of the Multiscale Statistic} According to (\ref{equ:mutiscale_statistic}), the multiscale statistic includes dissimilarity statistic for lumped variables, spatial statistic, and temporal statistic for distributed variables. The dissimilarity statistic reflects the disorder degree of $ P $ numbers of terminal voltages measured from $ P $ parallel circuits. The inconsistent change of voltages will increase the disorder degree, thereby resulting in a relatively large dissimilarity statistic.

On the other hand, greater values of the spatial statistic or temporal statistic indicate an abnormal temperature distribution in the spatial domain or abnormal temperature fluctuations in the temporal domain. These abnormal distributions and variations can be regarded as the manifestations of the increase in system disorder degree. Therefore, the comprehensive multiscale statistic can detect the inconsistent change of terminal voltages, the abnormal temperature distribution, and the abnormal variation of cell temperatures.


\subsection{MIF-based Abnormality Detection and Localization}
Substituting the dissimilarity statistic (\ref{equ:dissimilarity_statistic}), the spatial statistic (\ref{equ:spatial_statistic}), and the temporal statistic (\ref{equ:temporal_statistic}) into the optimization (\ref{equ:optimization_objective}), the optimal parameters of the multiscale statistic $ H(k) $ can be found. Based on the optimized $ H(k) $, the abnormality can be detected and identified as follows.   
\subsubsection{Threshold Design} In order to improve the ADR and reduce the FAR, a reference signal (threshold) of the multiscale statistic should be appropriately designed before abnormality detection and localization. The method of kernel density estimation (KDE), which is a non-parametric technique used to estimate the PDF of a random variable, is applied to approximate the PDF of the multiscale statistic as follows:
\begin{equation}\label{equ:kernel_density_estimation}
	g(\omega) = \frac{1}{bL_1} \sum_{k=1}^{L_1} K(\frac{\omega-H(k)}{b})
\end{equation}
where the bandwidth $ b $ can be calculated as $ b=1.06 \sigma_H L_1^{-1/5} $ with $\sigma_H$ denoting the sample standard deviation of the multiscale statistic. Here, the kernel function is selected as the Gaussian kernel \cite{sheather2004density}. Then, the reference signal $ H_r $ will be derived as:
\begin{equation}\label{key}
	\beta = \int_{0}^{H_r} g(\omega) d\omega
\end{equation}
where $ \beta $ denotes the confidence level.

\subsubsection{Abnormality Detection} After the design of the reference signal $ H_r $ of the multiscale statistic, the implementation of abnormality detection can proceed as follows:
\begin{itemize}
	\item [(1)] When $H(k)$ exceeds the threshold $H_r$, it signifies the detection of an abnormality, and the time at which it occurs is noted as $t_f$.
	\item [(2)] Otherwise, the system is deemed to be in a normal state.
\end{itemize}

\subsubsection{Abnormality Localization} After an abnormality is detected, the location of the abnormality can be identified as follows. First, the contribution function of the spatial statistic can be constructed as:
\begin{equation}\label{key}
	C(z) = \frac{1}{nW}\sum_{k=t_f-W+1}^{t_f} \sum_{i=1}^{n} |\phi_i^k - \phi_i^0|
\end{equation}
where $ z $ is the spatial location and defined as $ z=[x,y]^T $. Then, the coordinate corresponding to the maximum of $ C(z) $ can be regarded as the abnormality position, i.e., 
\begin{equation}\label{key}
	z_f = \arg \max_z \frac{1}{nW}\sum_{k=t_f-W+1}^{t_f} \sum_{i=1}^{n} |\phi_i^k(z) - \phi_i^0(z)|
\end{equation}
in which $ z_f $ represents the coordinate of the abnormality.

\section{Experimental verification}
\subsection{Experimental Configuration}
\begin{figure}[htbp] 
	\centering
	\includegraphics[width=0.35\textwidth]{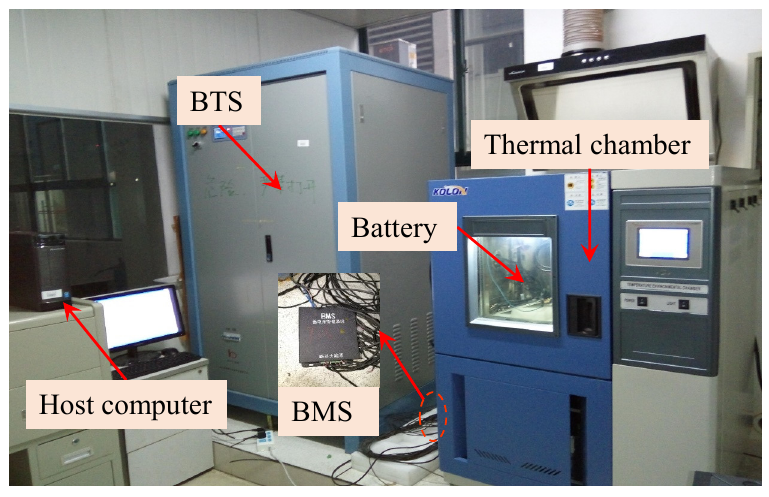}
	\caption{Experimental test bench.}
	\label{fig:test_bench}	
\end{figure}

\begin{figure}[htbp] 
	\centering
	\includegraphics[width=0.30\textwidth]{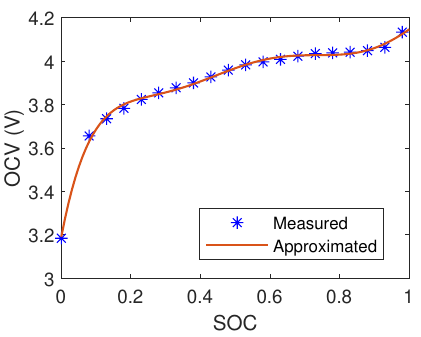}
	\caption{Measured points and identified OCV-SOC curve.}
	\label{fig:OCV_SOC}	
\end{figure}

In order to accurately replicate real-world conditions, the approach for parameter identification method described in Ref. \cite{lao2018novel} is utilized to establish the relationship between the open-circuit voltage (OCV) and the state of charge (SOC) of the battery cell with the experimental test bench shown in Fig. \ref{fig:test_bench}. The test bench is employed to collect experimental data of battery cells during charging and discharging. The identified OCV-SOC curve is shown in Fig. \ref{fig:OCV_SOC}, and the function is as follows: 
\begin{equation}\label{}
	\begin{split}
		OCV = &-34.39 \times SOC^6 + 127.38 \times SOC^5 \\
		&-182.10 \times SOC^4 + 127.24 \times SOC^3  \\
		&-45.57 \times SOC^2 + 8.40 \times SOC	+3.19
	\end{split}
\end{equation}

The experimental test bench comprises a host computer, a battery test system (BTS), a thermal chamber, and a battery management system (BMS). The BTS module is capable of generating various current waveforms for both charging and discharging the battery in accordance with control signals from the host computer. The thermal chamber is employed to regulate the ambient temperature during battery testing. The BMS module is responsible for gathering experimental data, including voltage, current, and temperature, and then transmitting this data to the host computer for further analysis and processing.

\begin{figure}[htbp] 
	\centering
	\includegraphics[width=0.30\textwidth]{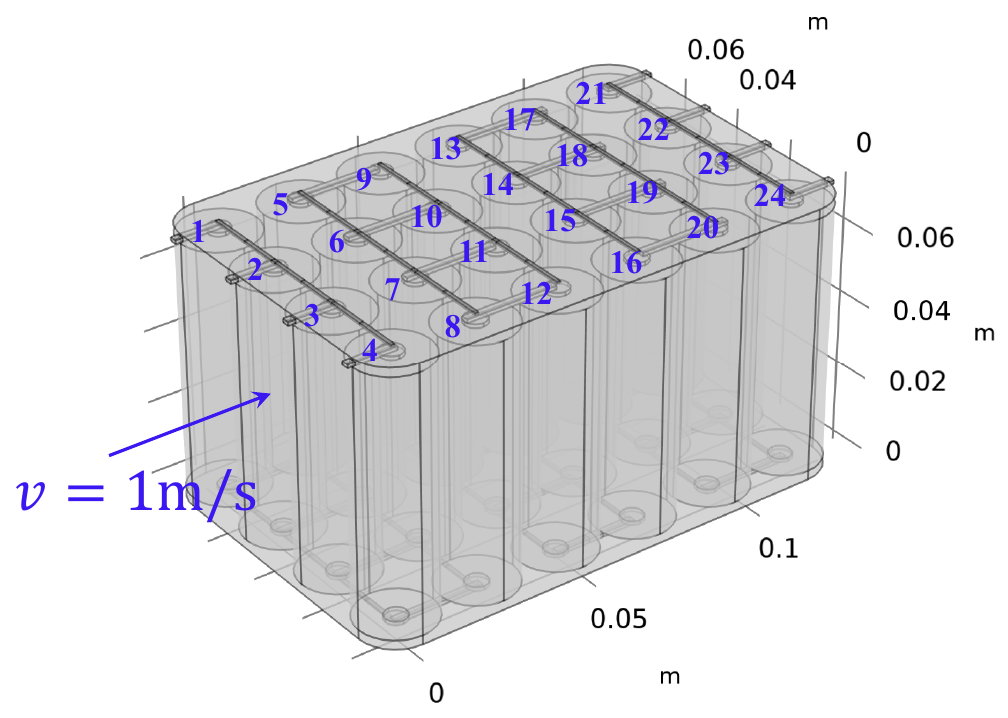}
	\caption{3D sketch of the battery system.}
	\label{fig:battery_pack}	
\end{figure}

Since thermal failure experiments are expensive, hazardous, and challenging to replicate with real-time data, the modeling approach outlined in \cite{saw2016computational} is employed to simulate both normal and abnormal scenarios in the battery pack. As depicted in Fig. \ref{fig:battery_pack}, the battery system comprises 24 cylindrical battery cells and is employed for evaluating the performance of the proposed method. Initially, four cells are connected in parallel, followed by the connection of six ($P=6$) parallel pairs in series. Each battery cell is labeled with its serial number on the top. To replicate the actual cooling process, there is directional airflow ($v=1$\ m/s) on the left side of the battery system. The key model parameters and their sources are detailed in Table \ref{tab:parameters}.

\begin{table}[htbp] 
	\centering
	\begin{threeparttable}
		\caption{Parameter definitions and sources}
		\label{tab:parameters}
		\begin{tabular} {cccc} 
			\toprule 
			Parameter  &  Value & Unit & Source \\
			\midrule 
			Diameter of battery cell  &     0.021  & m & Measured \\
			Height of battery cell &     0.070  & m & Measured \\			
			Nominal capacity of battery cell          &   4.8    & Ah & Handbook\\
			Nominal voltage of battery cell        &   3.7    & V & Handbook\\
			Gap between outer walls &     0.002  & m & Selected\\
			Quantity of cells within the pack             &  24   &    -   &   Selected\ \\
			Number of sensors $ N $    &     24       & - & Selected \\
			Reference time $t_r$ & 1000 & s & Selected \\
			Ambient temperature        &  293.15    & K & Selected\\
			Window size $ W $    &     27       & - & Optimized \\
			Weighting parameter $ \alpha_1 $	       & 	0.216    & -	& Optimized \\
			Weighting parameter $ \alpha_2 $	       & 	0.573    & -	& Optimized \\
			Weighting parameter $ \alpha_3 $	       & 	0.211    & -	& Optimized \\
			Confidence level $ \beta $	       & 	 0.99   & -	& Selected \\
			Model order $ n $	       & 	 5   & -	& Derived \\
			\bottomrule
		\end{tabular}
	\end{threeparttable}
\end{table}

An internal short circuit (ISC) within a battery cell is employed to simulate thermal abnormalities in the battery system. In the early stages, the thermal process of a small resistance can be considered as the effect of ISC on the battery system, as discussed in \cite{feng2016online}. The corresponding power density function, denoted as $P(t)$, can be calculated using the following expression:
\begin{equation}\label{key}
	P(t) = \frac{V^2}{R_{\text{short}}}\bigg/\left(\frac{4}{3}\pi r^3\right) = \frac{3V^2}{4\pi r^3 R_{\text{short}}}
\end{equation}
where $V$ denotes the terminal voltage, while $R_{\text{short}}$ denotes the equivalent ISC resistance value. $r$ signifies the equivalent radius of the ISC resistor. For the purposes of this research, $r$ has been assigned a value of 0.005 m. The detailed abnormality settings are listed in Table \ref{tab:settings_of_faults}. In each experiment, 2,000 data sets are collected, of which the first 600 are used for reference signal calculation, and the last 1,400 are used for testing.

\begin{table}[htbp]
	\centering\small
	\begin{threeparttable}
		\caption{Configuration of ISC Abnormalities}
		\label{tab:settings_of_faults}
		\begin{tabular} {ccccc} 
			\toprule 
			No.& Position & \makecell{Discharge\\ rate} &  \makecell{$R_\text{short}$\\$(\Omega)$} &\makecell{Occurring \\time (s)}\\
			\midrule 
			1                & \#4    & 2 &  10 & 1000 \\
			2                & \#5    & 2 &  10 & 1000 \\
			3                & \#11    & 2 &  10  & 1000 \\
			4                & \#16    & 2 &  10 & 1000 \\
			5                & \#18    & 2 &  10 & 1000 \\
			6                & \#23    & 2 &  10  & 1000 \\
			7                & \#23    & 1 &  10  & 1000 \\
			8                & \#23    & 2 &  5  & 1000 \\
			9                & \#23    & 2 &  10  & 1500 \\
			\bottomrule 
		\end{tabular}
	\end{threeparttable}
\end{table}	

\subsection{Abnormality Detection}
\begin{figure}[htbp]
	\centering
	\subfigure[Proposed multiscale statistic $ H(t) $.]{
		\includegraphics[width=0.23\textwidth]{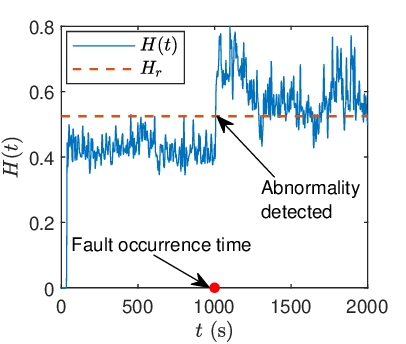}}
	\subfigure[$ SPE(t) $ of PCA.]{
		\includegraphics[width=0.23\textwidth]{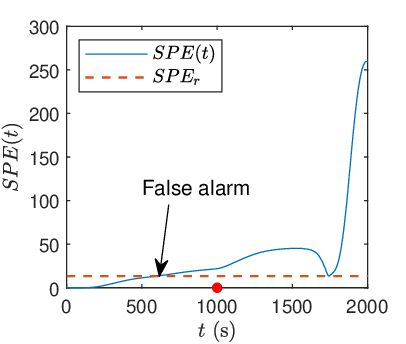}}
	\caption{Comparative analysis of abnormality detection on Fault 1}
	\label{fig:detection_fault1}
\end{figure}

Fig. \ref{fig:detection_fault1} (a) illustrates the results of abnormality detection using the proposed method under Fault 1 conditions. The actual time of fault occurrence is marked by the solid red circle on the horizontal axis. The red dashed line marks the reference signal of the proposed $H(t)$ statistic. Abnormalities are detected when the statistic surpasses its reference signal. During the testing phase, the first instance when the proposed $ H(t) $ statistic exceeds its reference signal is marked with an arrow. As depicted in Fig. \ref{fig:detection_fault1} (a), shortly after the occurrence of ISC, the proposed $ H(t) $ statistic exceeded its reference signal, signifying the ability of the proposed method to swiftly detect abnormalities.

To make a comparison, the squared prediction error (SPE) statistic from the PCA method is utilized. The detection outcome of the SPE statistic under Fault 1 conditions is illustrated in Fig. \ref{fig:detection_fault1} (b). Before an abnormality occurs, the $ SPE(t) $ statistic has exceeded its reference signal for a long time. This shows that the ISC abnormalities are too small to be detected in the residual space, so the traditional $ SPE $ statistic is ineffective for abnormality detection of the battery system.
\begin{table}[htbp]
	\centering \small
	\begin{threeparttable}
		\caption{Results of Abnormality Detection and Localization of the Proposed Method}
		\label{tab:performance}
		\begin{tabular} {cccccc} 
			\toprule 
			\makecell{No.} & ADD (s) & ADR (\%)  & FAR (\%) & \makecell{Estimated\\fault cell} \\
			\midrule 
			1 &11 &92.20 & 0.20 & \#4 \\
			2 &15 &86.30 & 0.20 & \#5 \\
			3 &13 &90.70 & 0.30 & \#11 \\
			4 &15 &88.10 & 0.50 & \#16 \\
			5 &16 &85.50 & 0.30 & \#18 \\
			6 &12&86.80 & 0.20 & \#23 \\
			7 &19 &90.40 & 1.80 & \#23 \\
			8 &14 &93.70 & 1.90 & \#23 \\
			9 &20&85.20 & 0.20 & \#23 \\
			\bottomrule 
		\end{tabular}
	\end{threeparttable}
\end{table}

As indicated in Table \ref{tab:performance}, the performance of the proposed method is quantitatively assessed using the ADD, the ADR, and the FAR. All ADDs are below 20 seconds, and all FARs are under 2\%, demonstrating the method's ability to timely detect abnormalities in the battery system and its reliability. However, there is room for improvement in sensitivity since the ADRs for Faults 2, 5, 6, and 9 are below 90\%.

As shown in Table \ref{tab:performance_comparison_with_spatiotemporal_entropy}, the proposed multiscale statistic is compared with the distributed temperature-based method in \cite{wei2022spatiotemporal} and the voltage-based method \cite{schmid2020data} under different conditions. The voltage abnormality is defined as a 5\% voltage drop in the sixth parallel circuit from 1500 s, i.e., $  \widetilde V_6(t) = 0.95*V_6(t),\ t\geq1500 $. The definition of the temperature abnormality is the same as that of Fault 9 in Table \ref{tab:settings_of_faults}. 

According to the results in Table \ref{tab:performance_comparison_with_spatiotemporal_entropy}, our multiscale statistic demonstrates superior performance compared to the distributed temperature-based method \cite{wei2022spatiotemporal} under voltage abnormality and surpasses the voltage-based method \cite{schmid2020data} under temperature abnormality. When voltage abnormality and temperature abnormality occur simultaneously, our proposed method performs better than  both peer methods.
Overall, our proposed method exhibits greater robustness compared to single-scale methods across various fault conditions.

\begin{table}[htbp]
	\centering\small
	\begin{threeparttable}
		\caption{Performance Comparison under Different Conditions}
		\label{tab:performance_comparison_with_spatiotemporal_entropy}
		\begin{tabular} {cccccccc} 
			\toprule
			\multicolumn{2}{c}{Fault type}& \multicolumn{6}{c}{Comparison methods}  \\
			\multirow{2}*{Volt.} & \multirow{2}*{Temp.}
			&\multicolumn{2}{c}{Proposed} &\multicolumn{2}{c}{Temp.-based \cite{wei2022spatiotemporal}}
		    &\multicolumn{2}{c}{Volt.-based \cite{schmid2020data}}\\ 
			\cline{3-8}
			\rule{0pt}{12pt}  
			~&~& ADR & ADD& ADR& ADD& ADR& ADD  \\
			\midrule
			\XSolid &  \Checkmark  & 85.20& \textbf{20s} & \textbf{86.80}& \textbf{20s} &5.90 &137s\\
			\Checkmark  & \XSolid    & 74.40& \textbf{11s} &  6.80& 134s  &\textbf{76.30} & \textbf{11s}\\
			\Checkmark  & \Checkmark    & \textbf{88.60}& \textbf{8s} &  86.80& 20s & 76.30& 11s \\
			\bottomrule 
		\end{tabular}
		\begin{tablenotes}
			\item[] The best performance is marked in bold in the table.
		\end{tablenotes}
	\end{threeparttable}
\end{table}

\subsection{Abnormality Localization}
\begin{figure}[htbp]
	\centering
	\subfigure[$C(z)$ of Fault 1.]{
		\includegraphics[width=0.23\textwidth]{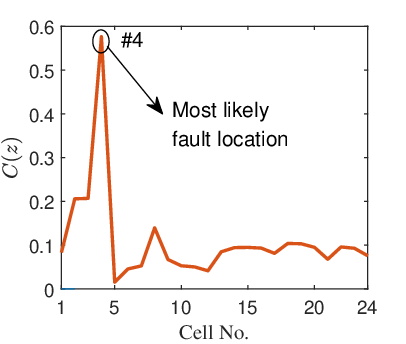}}
	\subfigure[$C(z)$ of Fault 9.]{
		\includegraphics[width=0.23\textwidth]{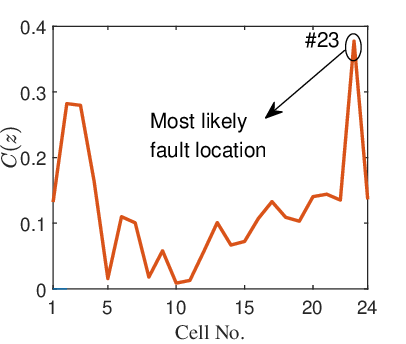}}
	\caption{Abnormality localization results of the proposed method.}
	\label{fig:localization_results}
\end{figure}

Figures \ref{fig:localization_results} (a) and (b) showcase the results of abnormality localization for Fault 1 and Fault 9, respectively. The maximum value of the spatial contribution function $ C(z) $, i.e., the most likely fault location, is marked with an arrow, and the corresponding cell serial number is listed next to it. The detailed localization results for all failure conditions are listed in the last column of Table \ref{tab:performance}. The estimated cell serial numbers match those in Table \ref{tab:settings_of_faults}, underscoring the ability of the proposed method to accurately pinpoint ISC abnormalities within the battery system.

\section{Conclusion}
The multiscale information fusion (MIF) has been proposed for abnormality detection and localization of DPSs under various scenarios. The proposed multiscale statistic has good physical interpretability and demonstrates more reliable performance than traditional single-scale detection methods under different working conditions. The experiments conducted on a battery pack have confirmed that the proposed method can effectively detect ISC abnormalities within 20 seconds while maintaining a low 2\% false alarming rate, and it can also accurately locate the abnormal battery cells. The proposed multiscale information fusion framework offers a theoretical foundation for the statistical design of reliable multiple-fault detection systems.

\bibliographystyle{IEEEtran}
\bibliography{my_ref7}

\begin{IEEEbiography}[{\includegraphics[width=1in,height=1.25in,clip,keepaspectratio]{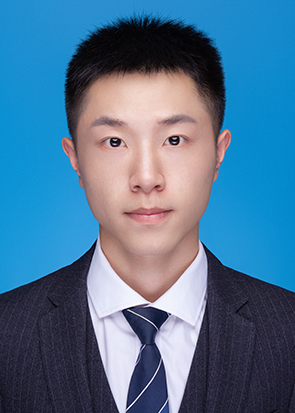}}]{Peng Wei}
	received the B.E. degree in mechanical engineering from Huazhong Agricultural University, Wuhan,
	China, in 2017, the M.E. degree in mechanical engineering from Huazhong University of Science and Technology, Wuhan, China, in 2019, and the Ph.D. degree in systems engineering from City University of Hong Kong, Hong Kong, China, in 2023.
	\par His current research interests include spatiotemporal modeling and fault diagnosis of distributed parameter systems, especially battery energy storage systems.
\end{IEEEbiography}

\begin{IEEEbiography}[{\includegraphics[width=1in,height=1.25in,clip,keepaspectratio]{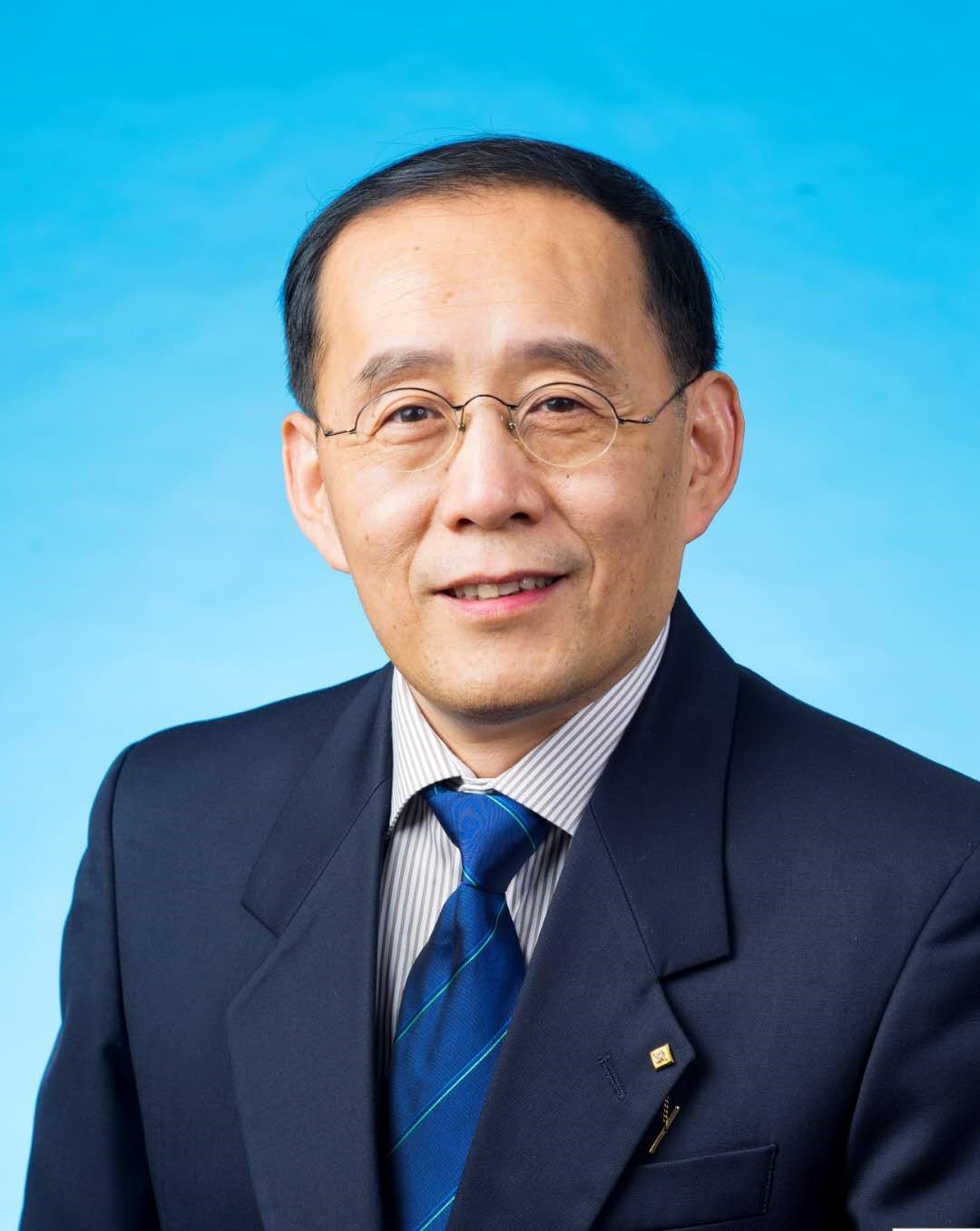}}]{Han-Xiong Li}
	(Fellow, IEEE) received the B.E. degree in aerospace engineering from the National University of Defense Technology, Changsha, China, in 1982, the M.E. degree in electrical engineering from the Delft	University of Technology, Delft, The Netherlands, in 1991, and the Ph.D. degree in electrical engineering from the University of Auckland, Auckland, New Zealand, in 1997.
	
	He is a Chair professor in the Department of Systems Engineering, City University of Hong Kong. He has broad experience in both academia and industry. He has authored 3 books and about 20 patents, and published more than 250 SCI journal papers with H-index 58 (web of science). His current research interests include process modeling and control, system intelligence, distributed parameter systems, and battery management systems.
	
	Dr. Li serves as Associate Editor for IEEE Transactions on SMC: System, and was associate editor for IEEE Transactions on Cybernetics (2002-2016) and IEEE Transactions on Industrial Electronics (2009-2015). He was awarded the Distinguished Young Scholar (overseas) by the China National Science Foundation in 2004, a Chang Jiang professorship by the Ministry of Education, China in 2006, and a national professorship in China Thousand Talents Program in 2010. He serves as a distinguished expert for Hunan Government and China Federation of Returned Overseas Chinese. He is a Fellow of the IEEE.
\end{IEEEbiography}

\end{document}